\documentclass[aps,prb,reprint,a4paper,twocolumn,showpacs,floatfix,citeautoscript]{revtex4-1}
\usepackage{graphicx}
\usepackage{amsmath,amssymb,mathrsfs}
\usepackage{epsfig}
\usepackage{hyperref}
\usepackage[usenames]{color}

\hypersetup{pdfborder=0 0 0,colorlinks=true,citecolor=blue,linkcolor=blue}

\begin{document}
\author{Yi Liu, Anton A. Starikov, Zhe Yuan, and Paul J. Kelly}
\affiliation{Faculty of Science and Technology and MESA$^+$ Institute for Nanotechnology, University of Twente, P.O. Box 217, 7500 AE Enschede, The Netherlands}
\title{First-principles calculations of magnetization relaxation in pure Fe, Co, and Ni with frozen thermal lattice disorder}
\date{\today}

\begin{abstract}
The effect of the electron-phonon interaction on magnetization relaxation is studied within the framework of first-principles scattering theory for Fe, Co, and Ni by displacing atoms in the scattering region randomly with a thermal distribution. This ``frozen thermal lattice disorder'' approach reproduces the non-monotonic damping behaviour observed in ferromagnetic resonance measurements and yields reasonable quantitative agreement between calculated and experimental values. It can be readily applied to alloys and easily extended by determining the atomic displacements from \textit{ab initio} phonon spectra.
\end{abstract}

\maketitle

\section{introduction}
The drive to increase magnetic storage densities and reduce access times is focussing renewed attention on magnetization dynamics in response to currents and external fields \cite{[See the collection of articles ]UMS}. The dynamics of a magnetization ${\mathbf M}$ in an effective field ${\mathbf H}_\textrm{eff}$ is usually described with the phenomenological Landau-Lifshitz-Gilbert equation
\begin{equation}
\frac{d{\mathbf M}}{dt}=-\gamma{\mathbf M}\times{\mathbf H}_\textrm{eff}+{\mathbf M}\times\left[\frac{\lambda({\mathbf M})}{\gamma M_s^2}\frac{d{\mathbf M}}{dt}\right]\,,
\end{equation}
where $M_s=\vert{\mathbf M}\vert$ is the saturation magnetization density and $\gamma=g\mu_0\mu_\mathrm{B}/\hbar$ the gyromagnetic ratio expressed in terms of the Land\'e $g$ factor and the Bohr magneton $\mu_\mathrm{B}$. The first term describes the precessional motion of the magnetization in the effective field that includes the external applied field, the exchange field, anisotropy and demagnetization fields. The second term describes the time decay of the magnetization precession, the Gilbert damping \cite{Gilbert:pr55,*Gilbert:ieeem04}, in terms of $\lambda({\mathbf M})$ that is in general a symmetric $3\times3$ tensor \cite{Steiauf:prb05}. For isotropic media, the damping is frequently expressed in terms of the dimensionless parameter $\alpha$ given by the diagonal element of $\lambda$, $\alpha=\lambda/\gamma M_s$.

There is general agreement that spin-orbit coupling (SOC) and disorder are essential ingredients in any description of how spin excitations relax to the ground state. In the absence of any other form of disorder, one might expect the damping to increase monotonically with temperature in clean magnetic materials and indeed, this is what is observed for Fe in ferromagnetic resonance (FMR) measurements \cite{Heinrich:pssb66,Bhagat:prb74}. Heinrich \textit{et al.} \cite{Heinrich:pssb67} developed an explicit model for this high-temperature behaviour in which itinerant $s$ electrons scatter from localized $d$ moments and transfer spin angular momentum to the lattice via SOC. This $s$-$d$ model results in a damping that is inversely proportional to the electronic relaxation time, $\alpha \sim 1/\tau$, i.e., is \textit{resistivity}-like. However, at low temperatures, both Co and Ni exhibit a sharp rise in damping as the temperature decreases \cite{Heinrich:jap79,Bhagat:prb74}. The so-called breathing Fermi surface model was proposed \cite{Kambersky:cjp70,Korenman:prb72,Kunes:prb02} to describe this low-temperature \textit{conductivity}-like damping, $\alpha\sim\tau$. In this model the electronic population lags behind the instantaneous equilibrium distribution due to the precessing magnetization and requires dissipation of energy and angular momentum to bring the system back to equilibrium. 

Of the numerous microscopic models that have been proposed \cite{Heinrich:05} to explain the damping behaviour of metals, only the so-called ``torque correlation model'' (TCM) \cite{Kambersky:cjp76} is qualitatively successful in explaining the non-monotonic behaviour observed for hcp Co. An effective field approach can be used \cite{Gilmore:prl07,*Gilmore:jap08,Kambersky:prb07,Gilmore:prb10} to identify conductivity-like and resistivity-like behaviour at low and high temperatures, respectively with intraband and interband terms in the TCM \cite{Heinrich:05}. Evaluation of this model for Fe, Co and Ni using first-principles calculations including SOC for the host electronic structure and a band-, wavevector- and spin-independent relaxation time approximation (RTA) to model disorder yields results for the damping $\alpha$ in good qualitative and reasonable quantitative agreement with the experimental observations \cite{Gilmore:prl07,*Gilmore:jap08,Kambersky:prb07,Gilmore:prb10}. The disadvantage of the RTA is that it is difficult to unambiguously map microscopically measured disorder onto a unique value of the relaxation time $\tau$. 

A formulation of magnetization damping in terms of scattering theory, that is equivalent in linear response to the Kubo formalism \cite{Brataas:prl08,*Brataas:arXiv11}, was recently applied to the study of substitutional alloys with intrinsic disorder yielding good agreement with experiment without introducing any parameters \cite{Starikov:prl10}. The discrepancies remaining between the experimental data measured at room temperature (and higher) and the $T=0$ calculations pose questions about the role of various types of thermal disorder. In this paper we combine the scattering theory formulation of Gilbert damping with structural lattice disorder to model finite temperature lattice effects. Our main result is to show that this can reproduce the non-monotonic behaviour of the magnetization relaxation as a function of temperature. 

The paper is organized as follows. In Sec.~\ref{sec:tmcd}, the ``frozen thermal lattice disorder'' scheme is introduced and a brief description of the scattering theory is given. This is followed by some technical details of how the calculations are performed and how the resistivity and Gilbert damping parameter are determined. The results are presented and discussed in Sec.~\ref{sec:rd} and compared to previous calculations and experiments. A short summary and some concluding remarks can be found in Sec.~\ref{sec:conc}. In the Appendix\ref{append}, we discuss a factor of $4\pi$ commonly omitted when Gilbert damping frequencies are given in Gaussian units.

\section{theoretical methods and computational details}
\label{sec:tmcd}
\begin{figure}
\includegraphics[width=0.9\columnwidth]{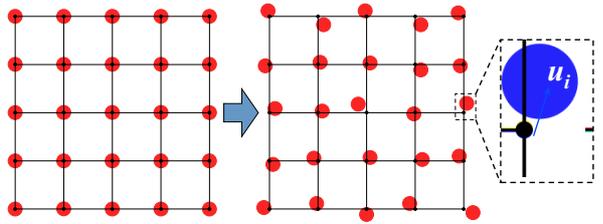}
\caption{(Color online) Schematic picture of the proposed frozen thermal lattice disorder. Atoms (red dots) on the ideal lattice (left panel) are displaced with a random Gaussian distribution and form a static configuration (right panel) in the electronic transport calculation. The displacement of atom $i$ is denoted as ${\mathbf u}_i$ (blow up).}\label{fig1}
\end{figure}

Assuming the Born-Oppenheimer approximation \cite{BO}, static disorder is introduced in the transport calculations by displacing atoms rigidly and randomly from their ideal lattice positions in what we call a ``frozen thermal lattice disorder'' scheme. In the temperature range we are interested in, far below the melting point, typical displacements are of the order of several hundredths of an angstrom, small compared to the lattice constant. We can therefore adopt a harmonic approximation and corresponding Gaussian distribution of displacements characterized by the root-mean-square (RMS) displacement, $\Delta=\sqrt{\langle\vert{\mathbf u}_i\vert^2\rangle}$ where the angular brackets indicate an average in which the index $i$ runs over all atoms (Fig.~\ref{fig1}). As the temperature increases, higher energy phonon modes are occupied so $\Delta$ increases. In the present study, this qualitative correlation between temperature and $\Delta$ is sufficient to produce a non-monotonic damping. If we knew the phonon dispersion relation from a Debye model or \textit{ab initio} calculations, the atomic displacements could be determined explicitly by summing contributions from all vibrational modes occupied at a specified temperature. Such a description of the lattice disorder introduced by finite temperatures could then be straightforwardly combined with scattering theory to study temperature dependent magnetization relaxation quantitatively.

It was shown by Brataas \textit{et al.}\cite{Brataas:prl08} that, for a single domain ferromagnetic metal (FM) sandwiched between left- and right-hand leads of non-magnetic (NM) material, the Gilbert damping tensor $\tilde{G}$ can be expressed as
\begin{equation}
\tilde{G}_{i,j}({\mathbf m})=\lambda (\mathbf m)\cdot V=\frac{\gamma^2\hbar}{4\pi}\textrm{Re}\left\{\textrm{Tr}\left[\frac{\partial S}{\partial m_i}\,\frac{\partial S^\dagger}{\partial m_j}\right]\right\}\,,\label{eq:gilbert}
\end{equation}
where $V$ is the volume of the ferromagnet, the scattering matrix 
$S=\left(\begin{array}{cc}r&t'\\t&r'\end{array}\right)$ is given in terms of reflection and transmission matrices for Bloch waves incident from the left ($r$ and $t$) or right ($r'$ and $t'$) leads. When SOC is included, $S$ depends on the direction of the magnetization unit vector ${\mathbf m}={\mathbf M}/M_s$. The microscopic picture of magnetization damping implicit in the scattering formulation is of energy being transferred slowly from the spin degrees of freedom through disorder scattering and SOC to the electronic orbital degrees of freedom and then being rapidly lost to phonon degrees of freedom in thermal reservoirs attached to the leads. From the transmission matrices, we can also calculate the conductance of the system within the Landauer-B{\"u}ttiker formulation as $ G = (e^2/h) {\rm Tr} \left\{ tt^\dagger \right\} $.

To evaluate the scattering matrix at the Fermi level, we use a ``wave-function matching'' scheme \cite{Xia:prb01,*Xia:prb06} implemented with tight-binding linearized muffin-tin orbitals (TB-LMTOs)\cite{Andersen:prb75,*Andersen:prb86} that was recently extended to include SOC \cite{Starikov:prl10}. The electronic structure of the NM$|$FM$|$NM sandwich is first determined self-consistently using a surface Green's function method\cite{Turek:97} with a minimal basis of TB-LMTOs in the atomic sphere approximation. In the current study of Fe, Co and Ni, we consider Au$\vert$Fe$\vert$Au, Cu$\vert$Co$\vert$Cu and Cu$\vert$Ni$\vert$Cu sandwiches so the lattice constants of the NM leads and FM scattering regions match almost perfectly.\cite{LCs} The two-dimensional (2D) Brillouin zone (BZ) of the $1\times1$ unit cell is sampled with a $120\times120$ grid in the self-consistent calculations. 

Disorder is introduced by randomly displacing atomic spheres in the FM scattering region using the frozen thermal lattice disorder scheme described above. The calculations are rendered tractable by imposing periodic boundary conditions transverse to the transport direction. It turns out that good results can be achieved even when these so-called ``lateral supercells'' are quite modest in size. In practice, a $4\times4$ lateral supercell and a $28\times28$ 2D BZ grid were found to be sufficient for Fe, and a $5\times5$ supercell and a $32\times32$ grid for Co and Ni, respectively.\cite{2Dk} The thickness of the FM region ranges from 20 to 340 atomic monolayers. For every thickness of the ferromagnet, we average over a number of random disorder configurations. The sample-to-sample spread is small for large values of $\Delta$ and five configurations are sufficient; for small values of $\Delta$, as many as 35 configurations are used.

\begin{figure}
\includegraphics[width=0.9\columnwidth]{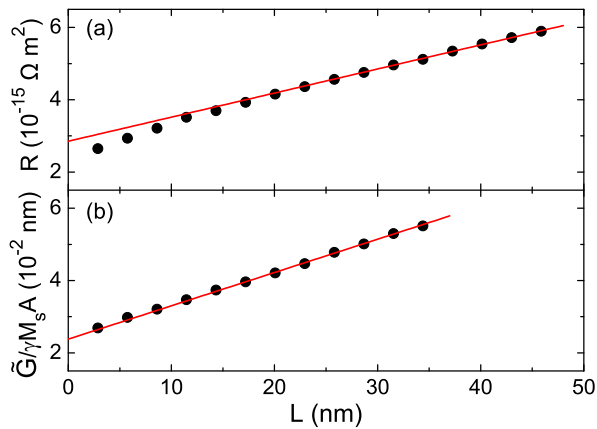}
\caption{(Color online) Total resistance (a) and damping (b) as functions of the thickness of the ferromagnetic slab for Au$\vert$Fe$\vert$Au(001) with $\Delta=0.0259\,a_\textrm{Fe}$. Dots indicate the calculated values averaged over five configurations while the solid line is a linear fit.}
\label{fit}
\end{figure}

For a ferromagnetic slab of thickness $L$, we write the resistance of the system as $R(L)=1/G(L)= 1/G_{\rm Sh}+2R_{\rm if}+R_{\rm b}(L)$, where $G(L)$ is the total conductance, and $G_{\rm Sh}$ the Sharvin conductance of the ideal leads; $R_{\rm if}$ is the NM$|$FM interface resistance and $R_{\rm b}(L)$ the bulk contribution \cite{Xia:prb01,*Xia:prb06,Schep:prb97}. When the ferromagnetic slab is sufficiently thick, we expect to recover ohmic behaviour with $R_{\rm b}(L)\approx\rho L$; this was demonstrated explicitly for the case of alloy disorder in Ref.~\onlinecite{Starikov:prl10}. As shown in Fig.~\ref{fit}(a), this expectation is borne out by the present calculations and the resistivity $\rho$ arising from the frozen thermal lattice disorder can be extracted from linear fitting. We write the damping parameter $\tilde{G}$ analogously as a sum of an interface contribution $\tilde G_\textrm{if}$ and a bulk contribution, $\tilde G_{\rm b}(L)$. If we further express the bulk contribution in terms of the dimensionless damping parameter $\alpha$ as $\tilde G_{\rm b}(L) = \lambda\cdot V= \alpha \gamma M_s AL$ where $A$ is the cross section, then we expect the calculated damping to grow linearly with the thickness of the ferromagnetic layers, $\tilde G(L)=\tilde G_{\rm if} + \alpha \gamma M_s AL$. Once again, this expectation is borne out by the calculations as demonstrated in Fig.~\ref{fit}(b), and $\alpha$ can be determined from the slope.

\begin{figure}
\includegraphics[width=\columnwidth]{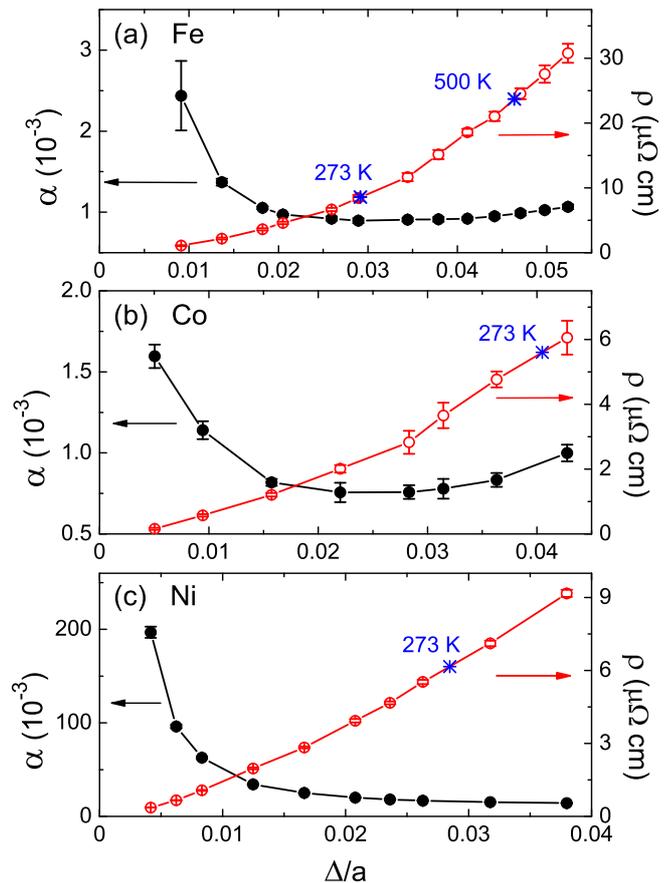}
\caption{(Color online) Calculated Gilbert damping and resistivity for bcc Fe, hcp Co and fcc Ni as functions of the RMS displacements measured in units of the corresponding lattice constants, $a$. The error bars reflect the configuration spread. Experimental resistivities \cite{HCP84} are used to label a number of resistivity values with a temperature.}\label{fig3}
\end{figure}

\section{results and discussion}
\label{sec:rd}

The resistivities and damping parameters calculated for bulk Fe, Co and Ni are shown as a function of the RMS displacement in Fig.~\ref{fig3}. For all three materials, $\rho$ increases monotonically with $\Delta$ as expected and $\alpha$ agrees with the predictions of the torque correlation model \cite{Kambersky:prb07,Gilmore:prl07,*Gilmore:jap08,Gilmore:prb10}. For large values of $\Delta$, $\alpha$ calculated for Fe and Co is found to increase with increasing $\Delta$, while tending to saturate for Ni in agreement with experiment \cite{Bhagat:prb74,Heinrich:pssb66,Heinrich:jap79}. For small values of $\Delta$, $\alpha$ increases rapidly as $\Delta$ decreases. This sharp rise is the conductivity-like behaviour observed at low temperature for Co and Ni, demonstrating that a simple model of frozen thermal lattice disorder can reproduce the non-monotonic Gilbert damping seen in experiment. 

To the best of our knowledge, this work represents the first calculation of $\rho$ for these ferromagnetic metals, only nonmagnetic materials having been considered in previous first-principles studies of temperature dependent resistivity \cite{Savrasov:prb96b}. While there is order-of-magnitude agreement with experiment, we cannot make a rigorous comparison if we only have a qualitative knowledge of how $\Delta$ depends on temperature $T$. If we adopt the commonly made assumption that the effect of temperature on $\rho$ and $\alpha$ can be expressed in terms of a phenomenological scattering time $\tau(T)$, i.e. $\rho(T)=\rho(\tau(T))$ and $\alpha(T)=\alpha(\tau(T))$, then $\alpha$ should have a well-defined dependence on $\rho$. In the present case, this amounts to assuming that our lattice disorder can be mapped onto $\tau(T)=\tau(\Delta(T))$. We can make a first qualitative comparison by using the experimental $\rho(T)$ \cite{HCP84} to indicate a number of  temperatures in Fig.~\ref{fig3}. Around these temperatures, the corresponding damping parameters behave in the same way as observed in experiment \cite{Bhagat:prb74,Heinrich:pssb66,Heinrich:jap79}: the damping for Fe has a minimum around 273~K and increases slightly at about 500~K; at 273~K, the damping for Co has increased away from its minimum value; while that for Ni has already saturated. This qualitative agreement indicates that the correlation between $\rho$ and $\alpha$ is reasonably parameterized by $\Delta$. In Table~\ref{table} we compare the minimum values of damping calculated as a function of $\Delta$ with the corresponding minimum values calculated within the TCM as a function of the relaxation time $\tau$ reported in Ref.~\onlinecite{Gilmore:prl07} and also with the minimum values determined experimentally as a function of temperature \cite{Bhagat:prb74,Heinrich:pssb66,Heinrich:jap79}. We see that our minimum values are lower than the TCM values that are, in turn lower than the experimental values.

There are two noteworthy qualitative differences between our results for small values of $\Delta$ and the low-temperature experimental observations. (i) For Fe, we find a conductivity-like damping behaviour that has not been seen in experiment \cite{Bhagat:prb74,Heinrich:pssb66} but has been found in the TCM calculations \cite{Kambersky:prb07,Gilmore:prl07,Gilmore:prb10}. (ii) For Ni and Co, the saturation of the damping observed at very low temperatures \cite{Bhagat:prb74,Heinrich:jap79} is found neither in our calculations nor in the TCM studies \cite{Kambersky:prb07,Gilmore:prl07,Gilmore:prb10}. We suggest that these discrepancies are in part related to extrinsic contributions to the Gilbert damping that compete with the intrinsic thermal effect. There will always be some amount of extrinsic disorder such as impurities and defects in experimental samples whose contribution to damping is essentially temperature independent and becomes dominant when the thermal disorder becomes negligible at sufficiently low temperatures. If the extrinsic disorder in the Fe samples measured in Refs.~\onlinecite{Bhagat:prb74,Heinrich:pssb66} was so high that Fe was in the resistivity-like regime\cite{regime} for all temperatures, then reducing the extrinsic disorder would be expected to lower the damping. This is consistent with the minimum damping value we calculate being smaller than the experimental value, see Table~\ref{table}, since our calculated damping comes only from the electron-lattice scattering, without any extrinsic contributions taken into account. For Ni and Co that exhibit both conductivity-like and resistivity-like behaviour, the discrepancies cannot be explained away so simply. Better characterisation of the experimental samples as well as further theoretical study are required.

\begin{table}
\caption{Minimum values of the Gilbert damping $\lambda$ in units of $10^8$ s$^{-1}$ with respect to: temperature (experiments); relaxation time, $\tau$ (Torque-correlation model); and $\Delta$ (present work). Experimental damping frequencies have been multiplied by $4\pi$; see the Appendix for details.
}
\label{table}
\begin{ruledtabular}
\begin{tabular}{l c c c}
$\lambda$              & bcc Fe [001] & hcp Co [0001]  & fcc Ni [111] \\
\hline
Expt.                    & 8.8\footnotemark[1], 4.8\footnotemark[2]          
                         & 9\footnotemark[1]              
                         & 29\footnotemark[1], 28\footnotemark[3]       \\
TCM\cite{Gilmore:prl07}  & 5.4          & 3.7            & 21           \\
This work                & 3.9          & 2.3            & 20           \\
\end{tabular}
\end{ruledtabular}
\footnotemark[1]Ref.~\onlinecite{Bhagat:prb74}; 
\footnotemark[2]Ref.~\onlinecite{Heinrich:pssb66}; 
\footnotemark[3]Ref.~\onlinecite{Heinrich:jap79}.
\end{table}

\begin{figure}
\includegraphics[width=\columnwidth]{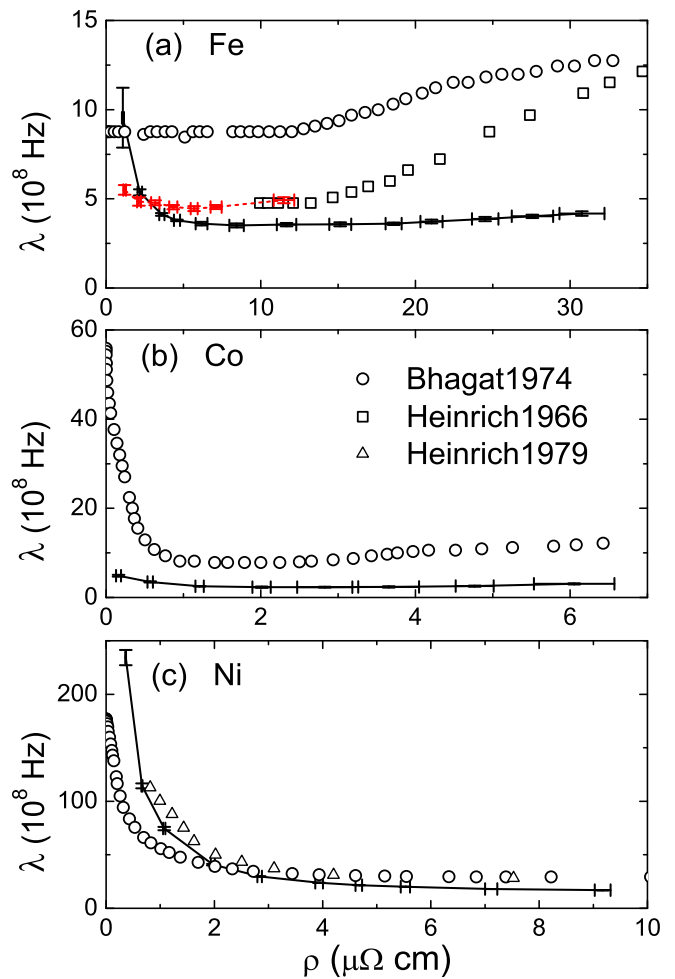}
\caption{(Color online) Gilbert damping frequency as a function of resistivity for bcc Fe, hcp Co and fcc Ni. Calculated results are shown as lines: for frozen thermal lattice disorder as (black) solid lines and  for frozen spin disorder as a (red) dashed line (Fe only). $g=2$ is used in the calculations. Symbols are experimental damping values from Ref.~\onlinecite{Bhagat:prb74} ($\bigcirc$), Ref.~\onlinecite{Heinrich:pssb66} ($\square$), and Ref.~\onlinecite{Heinrich:jap79} ($\triangle$). The corresponding resistivity values are take from References \onlinecite{HCP84} and \onlinecite{AIPhbk72}. Experimental damping frequencies have been multiplied by $4\pi$; see the Appendix \ref{append} for details.}
\label{fig4}
\end{figure}

Comparing the minimum values of $\lambda$, as we have just done, assumes that $\tau$ (or $\Delta$) characterizes $\lambda$ and $\rho$ completely. In particular, it assumes that the effects of various temperature-dependent disorder scattering mechanisms can be represented by a simple relaxation time. If this is correct, then the microscopic details of the disorder that give rise to particular values of $\rho$ and $\lambda$ should not be of paramount importance and we should be able to use experimentally determined $\lbrace \lambda(T), \rho(T) \rbrace$ and theoretically determined $\lbrace \lambda(\Delta), \rho(\Delta) \rbrace$ values to compare $\lambda_{\rm expt}(\rho)$ and $\lambda_{\rm calc}(\rho)$. Because we are not aware of any simultaneous measurements of $\lambda$ and $\rho$, we have combined $\lambda(T)$ and $\rho(T)$ from different experiments \cite{Bhagat:prb74,Heinrich:pssb66,Heinrich:jap79,HCP84,AIPhbk72} to plot $\lambda(\rho)$ in Fig.~\ref{fig4}; the calculated results are shown as black solid lines while the available experimental data are shown as symbols. To facilitate the comparison, the calculated damping parameter $\alpha$ in Fig.~\ref{fig3} is converted (see the Appendix\ref{append}) to damping frequency $\lambda$ using $\lambda=\alpha\gamma M_s$. For convenience we have used $g=2$ for the calculations instead of the measured $g$-factors.

The best agreement between the calculated curve and Bhagat and Lubitz's experimental values \cite{Bhagat:prb74} is obtained for Ni. As mentioned above, including the scattering associated with residual resistivity should {\em reduce} the damping calculated at low temperatures (resistivities) and increase it at high temperatures (resistivities) giving better agreement with experiment. However, it is not clear how the later measurements by Heinrich et al.\cite{Heinrich:jap79} might be accommodated in this picture. Clearly, there is a need to determined which experiment most accurately represents the ``intrinsic'' damping in Ni. A similar situation obtains for Fe where there is an even larger discrepancy between the two existing experiments. The figure highlights the disagreement between a conductivity-like behaviour at low values of resistivity in the calculations and its absence in both measurements. What is also striking is the failure of the calculations to reproduce the high temperature (resistivity) enhancement. For Co, where there is only one measurement, we find that in a reversal of the situation for Fe the conductivity-like behaviour is much more pronounced in experiment than in our calculations. It is unclear from this and the TCM studies how much of these discrepancies might result from subtle features of the electronic structure that are described inaccurately by the local spin density approximation. Tests with different exchange correlation potentials and increased maximum angular momentum cutoff indicate that the corresponding changes to the electronic structure at the Fermi energy cannot explain the discrepancies. 

We can test the uniqueness of $\tau$ directly by calculating $\lambda (\rho)$ for different types of disorder that give rise to the same resistivity. We do this for Fe by modelling frozen thermal spin disorder in a manner analogous to the way we have modelled frozen thermal lattice disorder. We introduce a random, Gaussian distribution of spins with respect to $\theta$, the polar angle between the local atomic magnetization and the global magnetization directions, together with a uniform random distribution in the azimuthal angle $\phi$. Together, $\theta$ and $\phi$ determine the local magnetization direction. We then calculate $\lambda$ and $\rho$ as a function of the root mean square polar rotation angle. The results are included in Fig.~\ref{fig4}a as a dashed (red) line. Though the qualitative behaviour of $\lambda_{\rm lattice}(\rho)$ and $\lambda_{\rm spin}(\rho)$ is the same, quantitatively they are clearly different, differing by as much as a factor of two for the lowest value of resistivity shown. This is clear direct evidence that different microscopic scattering mechanisms contribute to resistivity and Gilbert damping differently and that it may not be sufficient to use a single electronic scattering time to represent various types of disorder. It is also of interest to study to what extent the effect of different scattering mechanisms are additive. To do so, it is desirable to introduce ``real'' temperatures by calculating the phonon and magnon spectra from first-principles. This is a major undertaking and will be the subject of a separate study.

\section{Summary and Conclusions}
\label{sec:conc}
In summary, we report the results of first-principles calculations of the Gilbert damping with thermal lattice disorder for pure Fe, Co and Ni in the framework of a recently introduced scattering theory. The effect of temperature on the lattice is simulated by displacing atoms with a random, Gaussian distribution. Our main result is that both the  conductivity-like and resistivity-like damping behaviour observed in FMR measurements are reproduced by the scattering theory. The reasonable quantitative agreement between our results and experiment demonstrates that our simple thermal disorder scheme accounts for the dominant, intrinsic effect of lattice temperature in magnetization relaxation. By calculating the damping as a function of resistivity and comparing the results to experiment, we highlight discrepancies between different experiments and between the calculations and experiments. Part, but not all, of the discrepancies for the Gilbert damping can be attributed to competition between extrinsic and intrinsic scattering. An exploratory calculation of thermal spin disorder for Fe indicates that different types of disorder affect the Gilbert damping and resistivity in different manners. This work needs to be extended to different materials before more general conclusions can be drawn. It is of particular importance to study how different types of disorder combine to affect the damping. To avoid having to perform calculations in a two dimensional parameter space (lattice and spin disorder), it is desirable to introduce the effect of temperature using more realistic models for the lattice and spin disorder. The method we have used can be straightforwardly extended in this direction as well as to more complex materials. Finally, we hope that this work will stimulate more experimental temperature-dependent studies of magnetization damping.

\begin{acknowledgments}
We would like to thank Arne Brataas, Yaroslav Tserkovnyak and Gerrit Bauer for helpful discussions. This work is part of the research programmes of ``Stichting voor Fundamenteel Onderzoek der Materie'' (FOM) and the use of supercomputer facilities was sponsored by the ``Stichting Nationale Computer Faciliteiten'' (NCF), both financially supported by the ``Nederlandse Organisatie voor Wetenschappelijk Onderzoek'' (NWO). It was also supported by EU FP7 ICT Grant No. 251759 MACALO and Contract No. NMP3-SL-2009-233513 MONAMI.

\end{acknowledgments}

\appendix*
\section{$4\pi$ in Gilbert damping frequency}
\label{append}

The damping $\lambda$ has units of ${\rm s}^{-1}$ so its numerical value should not depend on whether SI or Gaussian units are used. Conversion of the dimensionless parameter $\alpha$ to the damping frequency $\lambda$ should follow the general relation
\begin{equation}
\lambda=\gamma\cdot\alpha\cdot\mathrm{magnetization~density}. 
\label{eq:convert}
\end{equation}
So converting from SI to Gaussian units should just require converting the magnetization density, $M_s$ (magnetic moment per unit volume that we calculate as $\mu_\mathrm{B}$/atom) from SI to Gaussian (cgs) units \cite{Jackson:99,Arrott:94}. In SI, $M_s$ is measured in units of A/m and the damping $\lambda$ is related to $\alpha$ by 
\begin{equation}
\lambda=\gamma\alpha M_s\,\mathrm{(SI)}.
\label{GSI}
\end{equation}
In Gaussian units, the magnetization should be measured in units of magnetic moment  cm$^{-3}$. If measured in Oersteds or Gauss, then the magnetization is $4\pi M_s$ and $\lambda$ is related to $\alpha$ by
\begin{equation}
\lambda=\gamma\cdot\alpha\cdot 4\pi M_s\, \mathrm{(in~Gaussian~units)}.
\label{Ggauss}
\end{equation}
In most experiments, the magnetization is measured and reported in Gaussian units but Eq.~\ref{GSI} is used instead of Eq.~\ref{Ggauss} which leads to a factor $4\pi$ missing in the damping frequency.

For example, a recent measurement\cite{Scheck:prl07} reports the magnetization of bcc iron to be $4\pi M_s=21.1$ kG, corresponding to $1.68\times 10^{6}$ A/m in SI or 2.13 $\mu_{\mathrm B}$/atom. Choosing $g=2$ (for simplicity), we have $\gamma=0.2203$ MHz$\cdot$m/A$= 17.588$ MHz/Oe. We convert the reported $\alpha=0.0019$ to $\lambda$  and obtain the same frequency $7.1\times 10^8$ Hz in SI using Eq.~\ref{GSI} or in Gaussian units using Eq.~\ref{Ggauss}. However, the frequency $57\pm 3$ MHz reported in Ref.~\onlinecite{Scheck:prl07}, which is obtained using Eq.~\ref{GSI} combined with magnetization in Gaussian units, has to be multiplied by $4\pi$ to be consistent.


\begin{thebibliography}{10}
\bibitem{BO}%
  \BibitemOpen
  \bibinfo {note} {In the scattering formulation the energy loss due to Gilbert
  damping is related to the energy pumped into leads by the precessing
  magnetization. The frequency corresponding to typical Fermi velocities in
  metals is $\sim 10^{16}$~Hz. Typical ferromagnetic resonance measurements
  from which the Gilbert damping constant is extracted are carried out at
  frequencies $\sim 10^{10}$~Hz so that transport electrons see essentially
  frozen spins. Spin-wave ($\sim 10^{13}$~Hz) and phonon frequencies ($\sim
  10^{12}$~Hz) are also much lower than the electronic frequencies, justifying
  the use of a frozen phonon approach.}%
  \bibAnnoteFile{Stop}{BO}%
\bibitem{LCs}%
  \BibitemOpen
  \bibinfo {note} {The lattice constant of bcc Fe is $a_\textrm{Fe}=2.867$~\AA,
  and that of fcc Au is chosen as
  $a_\textrm{Au}=\sqrt{2}\,a_\textrm{Fe}=4.054$~\AA. For Cu(111)$\vert$Co(0001)
  we use the experimental lattice constant $a_\textrm{Co}=2.507$ \AA, and a
  $c/a$ ratio of $1.623$. For Cu$\vert$Ni(111) $a_\textrm{Ni}=3.524$ \AA~is
  used. In both cases, the fcc Cu leads are slightly stretched to match.}%
  \bibAnnoteFile{Stop}{LCs}%
\bibitem{2Dk}%
  \BibitemOpen
  \bibinfo {note} {The 2D BZ areas for Co and Ni are almost identical and more
  than twice as large as that for Fe. Therefore the k-point sampling densities
  in the transport calculations for all three systems are very similar.}%
  \bibAnnoteFile{Stop}{2Dk}%
\bibitem{regime}%
  \BibitemOpen
  \bibinfo {note} {Resistivity-like and conductivity-like regimes are
  determined by whether additional disorder increases or decreases the
  damping.}%
  \bibAnnoteFile{Stop}{regime}%
\end{thebibliography}
\end{document}